\documentclass[journal]{IEEEtran}
\IEEEoverridecommandlockouts
\usepackage{cite}
\usepackage{amsmath,amssymb,amsfonts}
\usepackage{algorithmic}
\usepackage{graphicx}
\usepackage{textcomp}
\usepackage{xcolor}
\usepackage{subfigure}
\usepackage[colorlinks=true,
            linkcolor=blue,
            anchorcolor=blue,
            citecolor=blue]{hyperref}
\usepackage{indentfirst}
\usepackage{mathrsfs}
\newtheorem{remark}{Remark}
\usepackage{multirow}
\usepackage[ruled]{algorithm2e}
\usepackage{multirow} 
\usepackage{bm}
\usepackage{parskip}
\usepackage{url}
\usepackage{dsfont}
\usepackage{utfsym}
\usepackage{color}
\def\BibTeX{{\rm B\kern-.05em{\sc i\kerwn-.025em b}\kern-.08em
    T\kern-.1667em\lower.7ex\hbox{E}\kern-.125emX}}
    
\usepackage{booktabs}
\usepackage{array}
\newcolumntype{I}{!{\vrule width 1.05pt}}

\usepackage[acronym]{glossaries-extra}
 
\setabbreviationstyle[acronym]{long-short}
\newacronym{rsma}{RSMA}{Rate-Splitting Multiple Access}
\newacronym{ran}{RAN}{Radio Access Networks}
\newacronym{mec}{MEC}{Mobile Edge Computing}
\newacronym{mcc}{MCC}{Mobile Cloud Computing}
\newacronym{ao}{AO}{Alternative Optimization}
\newacronym{fbl}{FBL}{Finite Blocklength}
\newacronym{scp}{SCP}{Successful Computation Probability}
\newacronym{bcd}{BCD}{Block Coordination Descent}
\newacronym{sca}{SCA}{Successive Convex Approximation}
\newacronym{noma}{NOMA}{Non-orthogonal Multiple Access}
\newacronym{oma}{OMA}{Orthogonal Multiple Access}
\newacronym{tdma}{TDMA}{Time Division Multiple Access}
\newacronym{fdma}{FDMA}{Frequency Division Multiple Access}
\newacronym{ofdma}{OFDMA}{Orthogonal Frequency Division Multiple Access}
\newacronym{bs}{BS}{Base Station}
\newacronym{csi}{CSI}{Channel State Information}
\newacronym{csir}{CSIR}{Channel State Information at Receiver}
\newacronym{csit}{CSIT}{Channel State Information at Transmitter}
\newacronym{siso}{SISO}{Single-Input Single-Output}
\newacronym{awgn}{AWGN}{Additive White Gaussian Noise}
\newacronym{sic}{SIC}{Successive Interference Cancellation}
\newacronym{sinr}{SINR}{Signal to Interference plus Noise Ratio}
\newacronym{snr}{SNR}{Signal to Noise Ratio}
\newacronym{ifbl}{IFBL}{Infinite Blocklength}
\newacronym{ues}{UEs}{User Equipments}
\newacronym{mimo}{MIMO}{Multiple-Input Multiple-Output}
\newacronym{qos}{QoS}{Quality of Service}
\newacronym{5g}{5G}{5th Generation}
\newacronym{6g}{6G}{6th Generation}
\newacronym{sdma}{SDMA}{Space Division Multiple Access}
\newacronym{cpu}{CPU}{Central Processing Unit}
\newacronym{rhs}{RHS}{Right-Hand-Side}
\newacronym{lhs}{LHS}{Left-Hand-Side}
\newacronym{wrt}{w.r.t}{with respect to}

\title{Uplink Rate-Splitting Multiple Access for Mobile Edge Computing with Short-Packet Communications}

\begin{document}
\author{
\IEEEauthorblockN{Jiawei~Xu, ~Yumeng Zhang, ~Yunnuo Xu, ~Bruno~Clerckx, \IEEEmembership{Fellow, IEEE}}
\thanks{J. Xu is with Imperial College London. B. Clerckx is with the Department of Electrical and Electronic Engineering at Imperial College London, London SW7 2AZ, UK. (email: {j.xu20,b.clerckx}@imperial.ac.uk. Yumeng Zhang is with the Department of Electronics and Computer Engineering, The Hong Kong University of Science and Technology, Hong Kong 999077, China (e-mail: eeyzhang@ust.hk). Yunnuo Xu is with the School of Information Science and  Engineering, Shandong University, Qingdao 266237, China (e-mail:  yunnuo.xu@sdu.edu.cn)}
}
\maketitle

\begin{abstract}
In this paper, a Rate-Splitting Multiple Access (RSMA) scheme is proposed to assist a Mobile Edge Computing (MEC) system where local computation tasks from two users are offloaded to the MEC server, facilitated by uplink RSMA  for processing. The efficiency of the MEC service is hence primarily influenced by the RSMA-aided task offloading phase and the subsequent task computation phase, where reliable and low-latency communication is required. For this practical consideration, short-packet communication in the Finite Blocklength (FBL) regime is introduced. In this context, we propose a novel uplink RSMA-aided MEC framework and derive the overall Successful Computation Probability (SCP) with FBL consideration. To maximize the SCP of our proposed RSMA-aided MEC, we strategically optimize: (1) the task offloading factor which determines the number of tasks to be offloaded and processed by the MEC server; (2) the transmit power allocation between different RSMA streams; and (3) the task-splitting factor which decides how many tasks are allocated to splitting streams,  while adhering to FBL constraints. 
To address the strong coupling between these variables in the SCP expression, we apply the Alternative Optimization method, which formulates tractable subproblems to optimize each variable iteratively. The resultant non-convex subproblems are then tackled by Successive Convex Approximation. Numerical results demonstrate that applying uplink RSMA in the MEC system with FBL constraints can not only improve the SCP performance but also provide lower latency in comparison to conventional transmission scheme such as Non-orthogonal Multiple Access (NOMA).

\textit{Index Terms}\textemdash Rate-Splitting Multiple Access (RSMA), Non-orthogonal Multiple Access (NOMA), Mobile Edge Computing (MEC), Successful Computation Probability (SCP). 
\end{abstract}

\section{Introduction}\label{1}
It is envisioned that \gls{6g} will enable ubiquitous connectivity for a massive number of devices and provide low-latency and high-reliability communications services, such as the tactile Internet, remote surgery, and autonomous driving \cite{Li2019,8766143}. These computation-intensive and latency-sensitive applications challenge both the computational capabilities of \gls{ues} and the processing efficiency of \gls{mcc}. A significant bottleneck for \gls{mcc} lies in the long propagation distances between end users and remote cloud centers, resulting in substantial latency for mobile applications. It is widely acknowledged that relying only on Cloud Computing is insufficient to achieve the millisecond-level latency targets required for computing and communication in \gls{6g} networks \cite{8016573}. Additionally, the massive data exchange between end users and remote cloud servers risks overwhelming and congesting networks, potentially leading to network bottlenecks. In contrast to \gls{mcc}, \gls{mec} allows \gls{ues} to offload their workloads to edge servers for nearby processing, improving computing performance and lowering network latency \cite{7879258,abbas2017mobile}.

\par The concept of \gls{mec} was firstly proposed by the European
Telecommunications Standard Institute (ETSI) in 2014, and
was defined as a new platform providing IT and cloud computing capabilities within the \gls{ran} in close distance to mobile subscribers \cite{etsi}. The definition of \gls{mec} is based on offloading computation-intensive tasks from \gls{ues} to an edge device named \gls{mec} server which provides computing services to multiple \gls{ues}. Offloading data from a large number of \gls{ues} to \gls{mec} servers substantially exacerbates the issue of spectrum scarcity. Therefore, in order to achieve high spectral efficiency, energy efficiency and guarantee the \gls{qos} of \gls{mec} networks, wireless resources must be used effectively and appropriate multiple access techniques are needed. 
\cite{you2016energy,li2021energy,sun2019joint} have investigated the resource allocation problem in \gls{tdma}, \gls{fdma} and \gls{ofdma} aided-\gls{mec} systems, respectively, demonstrating the benefit of offloading. Compared to these \gls{oma} schemes, \gls{noma} has been recognized as an enabler to improve both the spectral and energy efficiency  \cite{saito2013non,islam2016power,khan2019joint,khan2020spectral}.

\par \gls{noma} has been widely investigated and applied in the \gls{mec} network to improve energy efficiency, reliability and/or latency \cite{8537962,8673584,zhu2020resource,9179779,8492422,ye2020enhance,9905567}. The authors combined \gls{noma} and \gls{mec} in \cite{8537962,8673584,zhu2020resource} to optimize the weighted sum-energy consumption with the constraints of computation latency. Offloading delay minimization problem of \gls{noma}-\gls{mec} network has been investigated in \cite{9179779,8492422}, where \cite{9179779} derived closed-form expressions for the optimal task partition ratio and offloading transmit power, \cite{8492422} has proposed and compared two algorithms to solve the optimization problem and further established the criteria to select the working modes of \gls{mec} offloading. 
As \gls{scp} which refers to the probability that a computation-intensive task is successfully completed, considering constraints such as latency and resources, is an important metric in an \gls{mec} system, an \gls{scp} maximization problem has been formulated and closed-form expressions for \gls{scp} have been derived in \cite{ye2020enhance}, demonstrating the gain of \gls{noma} over \gls{oma}. The overall error probability of \gls{noma}-aided \gls{mec} network with the consideration of imperfect \gls{sic} has been studied in \cite{9905567}. The numerical results have shown the performance improved with the implementation of \gls{noma}, however, multi-antenna \gls{noma} can impose heavy computational burdens on the transmitters and the receivers. Besides, it can also be sensitive to the user deployment. These disadvantages make \gls{noma} not suitable for \gls{6g} \gls{mec} network\cite{9831440}. 

\par In recent years, \gls{rsma}, relying on linearly precoded rate-splitting at the transmitter and \gls{sic} at the receiver, has emerged as a promising multiple access technique for \gls{6g}. By splitting user messages, \gls{rsma} can softly bridge \gls{noma} and \gls{sdma} \cite{Mao2018,10741240,9831048,clerckx2023primer}. Downlink and uplink \gls{rsma} have been demonstrated the capability of offering significant gains in terms of energy efficiency, user fairness, and latency reduction \cite{Mao2018,9970313,9991090,10741240,liu2023network} in both \gls{fbl} and \gls{ifbl}. There have been a few works investigating the application of \gls{rsma} in \gls{mec} network \cite{10243579,xiao2023delay,10032159}. The downlink \gls{rsma} principles have been utilized to facilitate the offloading of users' computation tasks to multiple \gls{mec} servers concurrently \cite{10243579}. A hybrid \gls{rsma}-\gls{tdma} scheme for an \gls{mec} system has been proposed to minimize delay \cite{xiao2023delay}. The sum of users' maximum delay has been minimized through jointly optimizing computation task assignment ratios, transmission power and computational resources. Instead of investigating delay minimization, the authors in \cite{10032159} have utilized \gls{rsma} in assisting \gls{mec} system to achieve the maximum achievable rate of secondary users while maintaining the performance of primary users. Besides, the authors have formulated an \gls{scp} maximization problem and derived the closed-form expressions for \gls{scp}. With the aid of \gls{rsma}, the performance of \gls{mec} network has been enhanced compared to \gls{noma}-aided \gls{mec} system \cite{ye2020enhance}. 

\par Although \gls{mec} has been widely investigated in terms of radio-and-computational resource allocation to improve the latency performance, aforementioned studies are based on conventional Shannon capacity, which is based on the assumption of an arbitrarily low decoding error probability with \gls{ifbl}, and it is no longer applicable to latency-critical systems. Because \gls{ifbl} does not fully reflect the real-world situation, where wireless communications are always subject to a certain reliability, and the transmission rate and the time slots allocation can impact the error probability of transmission \cite{8259329}. Polyanskiy et al. have provided information-theoretic limits on the achievable rate for given \gls{fbl} and error probability in \gls{awgn} channels \cite{5452208}. Then the maximum channel coding rate has been extended to quasi-static fading channels in \cite{7156144,ozcan2013throughput,xu2016energy} as well as in \gls{qos}-constraint network in \cite{hu2016blocklength} with constraints on error probability, blocklength and long-term transmit power for point-to-point communication scenarios. The maximum achievable transmission rate has been studied for quasi-static \gls{mimo} fading channels in \cite{yang2014quasi} under both \gls{csit}/\gls{csir} settings for a single user. To capture the impact of interference from multi-users within \gls{fbl} coding, \cite{scarlett2016dispersion,schiessl2018delay,hu2018optimal} have examined the achievable coding rate with a scenario involving $K$ users.

\par According to the latency-critical requirement of \gls{mec}, investigating \gls{mec} in \gls{fbl} regime becomes necessary. Some research has been carried out on \gls{mec} in \gls{fbl} regime \cite{zhu2022energy,9905567,lai2024short,liu2018offloading} in terms of energy efficiency, reliability, and latency, respectively. However, all those works are based on \gls{tdma} or \gls{noma} transmission schemes.The differences and comparisons between this work and previous works are listed in Table \ref{tab.1}. Motivated by 1) the need of state-of-art of multiple access in \gls{mec} network, 2) the appealing performance of \gls{rsma}, and 3) to capture the crucial feature of latency-critical requirements of \gls{mec}, we propose an uplink \gls{rsma}-assisted \gls{mec} framework and investigate its performance with the constraint of \gls{fbl} in terms of reliability and latency. The contributions of this work are summarized as follows.


\begin{table}[t!]
    \renewcommand{\arraystretch}{1.2}
    \caption{Comparison of this paper to the previous works.}
    \label{tab.1}
    \centering
    \begin{tabular}{IcIcIcIcIcI}
    \bottomrule[1.05pt]
    \hline
    Reference & \gls{noma} & \gls{rsma} & \gls{scp} & \gls{fbl} \\ \toprule[1.05pt]
		\bottomrule[1.05pt]
    \cite{you2016energy,li2021energy,sun2019joint} & $\scalebox{0.75}{\usym{2613}}$ & $\scalebox{0.75}{\usym{2613}}$ & $\scalebox{0.75}{\usym{2613}}$ & $\scalebox{0.75}{\usym{2613}}$ \\
    \hline
    \cite{9179779,8492422,9905567} & $\checkmark$ & $\scalebox{0.75}{\usym{2613}}$ &$\scalebox{0.75}{\usym{2613}}$ & $\scalebox{0.75}{\usym{2613}}$ \\
    \hline
    \cite{ye2020enhance} & $\checkmark$ & $\scalebox{0.75}{\usym{2613}}$ & $\checkmark$ & $\scalebox{0.75}{\usym{2613}}$ \\
    \hline
    \cite{10243579,xiao2023delay} & $\scalebox{0.75}{\usym{2613}}$ & $\checkmark$ & $\scalebox{0.75}{\usym{2613}}$ & $\scalebox{0.75}{\usym{2613}}$ \\
    \hline
    \cite{10032159} & $\checkmark$ & $\checkmark$ & $\checkmark$ & $\scalebox{0.75}{\usym{2613}}$ \\
    \hline
    \cite{zhu2022energy}  & $\checkmark$ & $\scalebox{0.75}{\usym{2613}}$ & $\scalebox{0.75}{\usym{2613}}$ & $\checkmark$ \\
    \hline
    \cite{9905567,lai2024short,liu2018offloading}  & $\checkmark$ & $\scalebox{0.75}{\usym{2613}}$ & $\checkmark$ & $\checkmark$ \\
    \hline
    \text{This work} & $\checkmark$ & $\checkmark$ & $\checkmark$ & $\checkmark$ \\\toprule[1.05pt]
\end{tabular}
\end{table}
\subsection{Contributions}
\begin{itemize}
    
    \item First, for the \textit{first time}, uplink \gls{rsma} with short-packet communications is introduced into \gls{mec} system as a powerful multiple access to facilate with the uplink task offloading procedure. The practical study of offloading task data with finite blocklength is noteworthy since it is an essential aspect for low-latency applications in \gls{6g} and future wireless networks.
    \item Second, the \gls{scp} performance of \gls{rsma}-aided \gls{mec} under \gls{fbl} constraints is analyzed. This work aims to maximize \gls{scp} performance by optimizing the entire system strategy with \gls{fbl} considerations, including the offloading factor, representing the number of tasks to be processed by the \gls{mec} server, the task-splitting factor, representing the number of offloading tasks assigned to each stream in the \gls{rsma} scheme and the power allocated to each splitting stream. A closed-form expression of the offloading factor is derived. Due to the coupling among the remaining optimization variables, \gls{ao} is applied to decompose the original problem into several sub-problems, each addressed by the proposed \gls{sca}-based method. This is the first work to offer a comprehensive optimization of systemic parameters for \gls{rsma}-aided \gls{mec} systems with \gls{fbl} constraints.  
    
    \item  Through simulations, the performance of \gls{rsma}-aided \gls{mec} system is analyzed. Simulation results show that splitting messages of \gls{rsma} helps in \gls{mec} deployments because of the flexibility of the decoding order to balance the \gls{sinr} between each stream.  Moreover, \gls{rsma} can achieve a higher \gls{scp} than \gls{noma} with a shorter blocklength, thereby reducing the latency. Besides, the performance gain of \gls{rsma} over \gls{noma} in the \gls{fbl} regime saturates as the blocklength becomes sufficiently large. Numerical results demonstrate \gls{rsma}-aided \gls{mec} achieves more reliable communication in \gls{fbl} regime with higher \gls{scp} compared to \gls{noma}.
\end{itemize}

\subsection{Organization}
The rest of the paper is organized as follows. The system model of the \gls{rsma}-aided \gls{mec}in with FBL is specified in Section \ref{2}. The \gls{scp} is defined and derived in Section \ref{3}. The problem is formulated and optimized in Section \ref{4}. Simulation results are presented in Section \ref{5} and we conclude the paper in Section \ref{6}. 

\subsection{Notations}
Italic and bold lower-case denote scalars and vectors, respectively. $|\cdot|$ denotes the absolute value if the argument is a scalar or the cardinality if the argument is a set. $\cdot\setminus\cdot$ denotes the difference between two sets if the arguments are sets. $\mathcal{CN}(\mu, \sigma^{2})$ denotes circular symmetric complex Gaussian distribution with mean $\mu$ and variance $\sigma^{2}$.

\section{System Model}\label{2}

This section models an \gls{rsma}-aided \gls{siso} \gls{mec} system with two users, indexed by $\mathcal{K}=\{1, 2\}$. Due to the limited local computation capabilities of the two users, this section characterizes an \gls{mec} scheme that schedules both users to simultaneously offload their data to an \gls{mec} server with high computation capability. Facilitated by the \gls{mec} server, both users can accomplish their tasks within the time budget by downloading the results from the \gls{mec} server. Specifically, it is assumed that each user, with a computational capability of $f_{\text{user}}$ Hz/cycle, needs to complete a computation task of $M_{k},~k\in\mathcal{K}$ bits within $T$ s. In practice, the scenario often arises that $M_{k}C_{\text{cpu}}/f_{\text{user}}>T$, where $C_{\text{cpu}}$ represents the number of required \gls{cpu} cycles to compute one-bit task, necessitating the involvement of a central \gls{mec} server with a computational capability of $f_{\text{MEC}}=Lf_{\text{user}}$ ($L > 1$). Assume that the $M_{k}$-bits task at each user is bit-wise independent and can be arbitrarily divided into subtasks. Hence, to meet the time budget $T$, each user-$k$ offloads a part of their task of $\lambda_{k}M_{k}$ bits, with $\lambda_{k}, ~0\leq \lambda_{k}\leq 1$, being the offloading factor of user-$k$, to the \gls{mec} server with the assistance of uplink \gls{rsma}. The remaining tasks of $(1-\lambda_{k})M_{k}$ bits are executed locally to exploit users' own local computation capability. $\lambda_{k}=0$ and $\lambda_{k}=1$ represent the fully local computation and fully offloading computation, respectively.

\begin{figure}[t]
    \centering
    \includegraphics[scale=0.55]{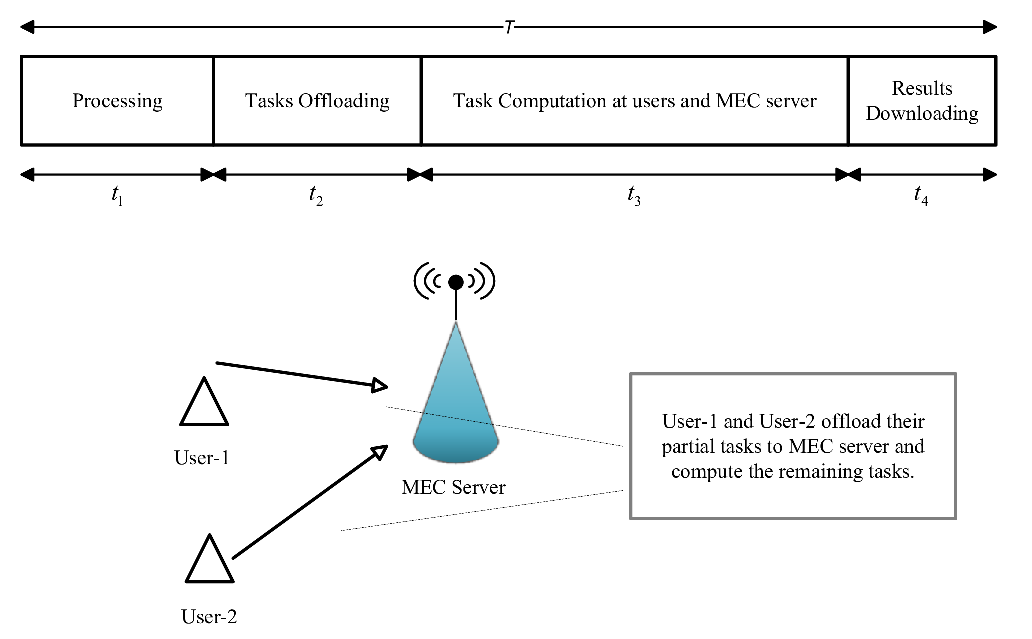}
    \caption{The \gls{rsma}-aided \gls{mec} system with two users.}
    \label{Fig.1}
\end{figure}

The whole procedure of the \gls{rsma}-aided \gls{mec} system is shown in Fig. \ref{Fig.1}, which consists of four stages. In the first stage, the offloading factor, task-splitting factor and power allocation are determined for two users within the processing time $t_{1}$ s. The rate-splitting power allocation factors guarantee the bit error rate performance of RSMA in the offloading stage. Then, in the second stage, users offload their tasks to the \gls{mec} server by uplink \gls{rsma} transmission within time duration $t_{2}$ s. Thereafter, the third stage performs task execution at both the \gls{mec} server and the two local users within time duration $t_{3}$ s. We assume the \gls{mec} server starts to compute the tasks immediately after the offloading is finished, i.e., there is no queue delay at the \gls{mec} server. In the last stage, the computed results of the offloaded tasks are downloaded by users, which consumes time $t_4$ s. Compared to the offloading time $t_{2}$ and execution time $t_{3}$, $t_{1}$ and $t_4$ can be ignored since the resultant data often features a very small size compared with the offloading data size in $t_{2}$ s \cite{wang2017joint}. Therefore, the time of offloading computation mainly comes from the time of tasks offloading stage $t_{2}$ and the time of tasks execution stage $t_{3}$, i.e., in the following of this paper, we set the time constraint to be $t_{2}+t_{3}\leq T$ without loss of generality.

In the reminder of this section, we detail the involved tasks offloading phase (stage 2) in subsection \ref{2.1}, and the execution parameters in tasks computation phase (stage 3) in subsection \ref{2.2}, respectively. Upon this, we are able to formulate the successful offloading probability of stage 2, the successful execution probability of stage 3 and the successful computation probability in Section \ref{3}.

\subsection{Tasks Offloading Phase}\label{2.1} 
In this subsection, the details of tasks offloading from two users to the \gls{mec} server are presented.

\subsubsection{RSMA}\label{2.A.1}
\par The two-user \gls{siso} uplink \gls{rsma} with perfect \gls{csit} and \gls{csir} system is considered. In the two-user uplink \gls{rsma}, one user's task is split following the uplink \gls{rsma} principle. We assume user-$1$ split its task $W_{1}$ into two parts, $W_{1,d}, ~d\in\{1, 2\}$ and the task of user-$2$ stays unsplit. These three split tasks $W_{1,1}$, $W_{1,2}$, and $W_{2}$ then are encoded independently, resulting into three streams $s_{1,1}$, $s_{1,2}$ and $s_{2}$, to be transmitted in total. The transmit power of user-$k$ is denoted as $P_{k}$. The received signal at the \gls{mec} server is expressed as
\begin{equation}\label{eq.1}
    y = h_1\sqrt{ P_{1,1}}s_{1,1}+h_1\sqrt{P_{1,2}}s_{1,2}+h_2\sqrt{P_{2}}s_2+n,
\end{equation}
where $P_{1,d}, ~d\in\{1,2\}$ is the transmit power of stream $s_{1,d}$ and $\sum_{d=1}^{2}P_{1,d}\leq P_{1}$. $n\sim\mathcal{CN}(0,\sigma_{n}^2)$ is the complex \gls{awgn} at the \gls{mec} server. 

\par At the \gls{mec} server, \gls{sic} is applied to decode three streams and all possible decoding orders can be classified into three cases, i.e., (\romannumeral1) $s_{1,1}\to s_{2}\to s_{1,2}$ (i.e. the \gls{mec} server decodes $s_{1,1}$ first, followed by $s_2$, and finally $s_{1,2}$), (\romannumeral2) $s_{1,1}\to s_{1,2}\to s_{2}$ and (\romannumeral3) $s_{2}\to s_{1,1}\to s_{1,2}$.\footnote{By exchanging the orders of $s_{1,1}$ and $s_{1,2}$, the other three decoding orders are ignored in this letter without loss of generality.} By utilizing chain rule, the achievable rates of user-$1$ and user-$2$ obtained by employing the decoding order (\romannumeral2) and (\romannumeral3) are equal to those obtained by employing the decoding order $s_{1}\to s_{2}$ and $s_{2}\to s_{1}$ of \gls{noma}, respectively. Besides, based on the previous study \cite{9970313}, uplink \gls{rsma} with the decoding order (\romannumeral1) can provide a much lower error probability compared to \gls{noma} in short-packet communication. Therefore, the decoding order $s_{1,1}\to s_{2}\to s_{1,2}$ is adopted in the following demonstration.

Since $s_{1,1}$ is decoded first by treating other signals as noise, the \gls{sinr} $\gamma_{1,1}$ of $s_{1,1}$ is 
\begin{equation}\label{eq.2}
    \gamma_{1,1} = \frac{P_{1,1}|h_{1}|^2}{P_{1,2}|h_{1}|^2+P_{2}|h_{2}|^2+\sigma_{n}^2}.
\end{equation}

Assuming $s_{1,1}$ is successfully decoded, its reconstructed task $\hat{W}_{1,1}$ is subtracted from the original received signal $y$ to obtain $y'$. Then, the \gls{sinr} of the second decoded stream $s_2$ expresses as 
\begin{equation}\label{eq.3}
    \gamma_{2} = \frac{P_{2}|h_{2}|^2}{P_{1,2}|h_{1}|^2+\sigma_{n}^2}.
\end{equation}

Once $s_{1,2}$ is successfully decoded, its reconstruction $\hat{W}_{1,1}$ is subtracted from the received signal $y'$. Thus, the \gls{sinr} of the last decoded stream $s_{1,2}$ is 
\begin{equation}\label{eq.4}
    \gamma_{1,2} = \frac{P_{1,2}|h_{1}|^2}{\sigma_{n}^2}.
\end{equation}

Finally, if $s_{1,2}$ is successfully decoded, the estimated task $\hat{W}_{1}$ for user-$1$ is obtained by combining $\hat{W}_{1,1}$ and $\hat{W}_{1,2}$. 

The data size of offloading tasks from three transmitted streams is defined as $M_{1,1}=\beta\lambda_{1}M_{1}$, $M_{2}=\lambda_{2}M_{2}$ and $M_{1,2}=(1-\beta)\lambda_{1}M_{1}$, where $\beta$ is the task-splitting factor satisfying $0\leq\beta\leq1$. Then, the total received tasks data size at the \gls{mec} server is given as
\begin{equation}\label{eq.5}
    M_{\text{MEC}}=\lambda_{1}M_{1}+\lambda_{2}M_{2}.
\end{equation}

Therefore, the Shannon capacity for each stream can be expressed as
\begin{equation}\label{eq.6}
    \begin{split}
        C(\gamma_{1,1}) & = \log_{2}(1+\gamma_{1,1}), \\
        C(\gamma_{2}) & = \log_{2}(1+\gamma_{2}), \\
        C(\gamma_{1,2}) & = \log_{2}(1+\gamma_{1,2}). 
    \end{split}
\end{equation}

Due to the consideration of uplink \gls{rsma} with \gls{fbl}, the total transmission time for offloading is 
\begin{equation}\label{eq.7}
    t_{2} = NT_{s},
\end{equation}
where $N$ is the blocklength (in symbol) and $T_{s}$ is the symbol duration. We assume the blocklength $N$ for each stream is the same.


\subsubsection{NOMA}
\par \gls{noma} is a particular instance of the system model of Section \ref{2.A.1}, since $0\leq P_{1,1} \leq P_{1}$, we can regard \gls{noma} as a subset of \gls{rsma}. The tasks $W_{1}$ and $W_{2}$ from user-1 and user-2 are encoded into $s_1$ and $s_2$, respectively. The received signal at the \gls{mec} server is expressed as $y=h_1\sqrt{P_1}s_1+h_2\sqrt{P_2}s_2+n$.
\begin{remark}
    If $P_{1,1}$ in \gls{rsma} scheme is set to $P_{1}$ or 0, \gls{rsma} boils down to \gls{noma} with decoding order of $s_{1}\to s_{2}$ or $s_{2}\to s_{1}$, respectively. Since $0\leq P_{1,1} \leq P_{1}$, we can regard \gls{noma} as a subset of \gls{rsma}.
\end{remark}

\subsection{Tasks Computation Phase}\label{2.2}
In this subsection, we discuss the details of the hybrid computations, including local computation at the users and offloading computation at the \gls{mec} server, respectively.

For the local computation at user-$k$, the data size of the tasks is $(1-\lambda_{k})M_{k}$, and the required computation time at user-$k$ is 
\begin{equation}\label{eq.8}
    t_{3}^{k} = \frac{(1-\lambda_{k})M_{k}C_{\text{cpu}}}{f_{\text{user}}},~\forall k\in\mathcal{K},
\end{equation}
where $C_{\text{cpu}}$ represents the number of CPU cycles required for one-bit task computing. $f_{\text{user}}$ denotes the computation capabilities of users and we assume the capacity of each user is the same.

For offloading computation at the \gls{mec} server, we assume the computation capability of the \gls{mec} server $f_{\text{MEC}}=Lf_{\text{user}}$ with $L> 1$ to indicate the difference between users' and \gls{mec} server's computational capabilities. Thus, the required computation time at the \gls{mec} server is 
\begin{equation}\label{eq.9}
    t_{3}^{\text{MEC}} = \frac{M_{\text{MEC}}C_{\text{cpu}}}{f_{\text{MEC}}} = \frac{M_{\text{MEC}}C_{\text{cpu}}}{Lf_{\text{user}}}.
\end{equation}

Thus, the time duration $t_{3}$ for tasks computation is expressed as
\begin{equation}\label{eq.10}
    \begin{split}
        t_{3} & = \max\{t_{3}^{k}, ~t_{3}^{\text{MEC}},~\forall k\in\mathcal{K}\} \\
        & = \max\{\frac{(1-\lambda_{k})M_{k}C_{\text{cpu}}}{f_{\text{user}}}, ~\frac{M_{\text{MEC}}C_{\text{cpu}}}{Lf_{\text{user}}},~\forall k\in\mathcal{K}\}.
    \end{split}
\end{equation}

\section{Successful Computation Probability Analysis}\label{3}

In this section, the \gls{scp} of \gls{rsma}-aided-\gls{mec} system with \gls{fbl} is characterized. The \gls{scp}, denoted by $P_{s}$, is defined as the probability of the case that all tasks (including offloaded tasks to be computed at the \gls{mec} server and the remaining tasks to be computed by the users) are computed successfully within the delay budget $T$. Thus $P_{s}$ contains two parts -- the successful offloading probability, $P_{s}^{o}$,  and the successful execution probability, $P_{s}^{e}$.

\subsection{Successful Offloading Probability}\label{3.1}
The Shannon capacity definition states that under the assumption of \gls{ifbl}, any rates inside the capacity region can be attained with an arbitrarily small error probability. However, in the case of an \gls{fbl} regime, there are chances that rates inside the capacity region cannot be obtained with an arbitrarily small error probability. 
Therefore, the maximal achievable rate expression for \gls{fbl} with a given error probability $\epsilon$, is given by \cite{5452208} 
\begin{equation}\label{eq.11}
    r \approx C(\gamma)-\sqrt{\frac{V(\gamma)}{N}}Q^{-1}(\epsilon),
\end{equation}
where $C(\gamma)$ is the Shannon capacity calculated by \eqref{eq.6} and $V(\gamma) = \biggl(1-(1+\gamma)^{-2}\biggl)$ is the channel dispersion. $N$ is the blocklength, and $Q$ is the Q-function \footnote{Q-function is the tail distribution function of the standard normal distribution. Normally, Q-function is defined as: Q(x)=$\frac{1}{\sqrt{2\pi}} \int_x^{\infty}e^{-\frac{u^2}{2}}du.$}. The error probability for a given data size $M$ can be derived based on \eqref{eq.11} as
\begin{equation}\label{eq.12}
    \epsilon = Q \biggl( \frac{\log_2(1+\gamma)-\frac{M}{N}}{\sqrt{V(\gamma)/N}}\biggl) = Q(f(\gamma, M)).
\end{equation}

For uplink \gls{rsma}, we can write the error probability for each event listed above as: a) $s_{1,1}$ is incorrectly decoded; b) $s_{1,1}$ is correctly decoded but $s_2$ is incorrectly decoded; c) $s_{1,1}$ and $s_2$ are both correctly decoded but $s_{1,2}$ is incorrectly decoded. The error probability of every event can be expressed as
\begin{equation}\label{eq.13}
    \begin{split}
        \epsilon_{a} & = Q(f(\gamma_{a}, M_{a})), \\
        \epsilon_{b} & = Q(f(\gamma_{b}, M_{b})), \\
        \epsilon_{c} & = Q(f(\gamma_{c}, M_{c})),
    \end{split}
\end{equation}
where $\gamma_{a}=\gamma_{1,1}$, $\gamma_{b}=\gamma_{2}$ and $\gamma_{c}=\gamma_{1,2}$. To simplify the analysis in the following Sec. \ref{4}, we change the subscript of $M_{k}$ and $\beta_{k}, ~k\in\mathcal{K}$ to align with the subscript notation of three splitting streams. Specifically, the parameter $M_{i}, ~i\in\{a,b,c\}$, represents the data size of tasks transmitted by each stream which are $M_{a}=\beta_{a}\lambda_{1} M_{1}$, $M_{b}=\beta_{b}\lambda_{2} M_{2}$ and $M_{c}=\beta_{c}\lambda_{1} M_{1}$, respectively. The parameter $\beta_{i},~i\in\{a,b,c\}$, represents the task-splitting factor between splitting streams where $\beta_{a}=\beta,~\beta_{b}=1$, and $\beta_{c}=1-\beta$. 
Then, the error probability of each user can be calculated as
\begin{equation}\label{eq.14}
\begin{split}
    \epsilon_1 & = \epsilon_a+(1-\epsilon_a)\epsilon_b+(1-\epsilon_a)(1-\epsilon_b)\epsilon_c \\
    & \approx  \epsilon_{a}+\epsilon_{b}+\epsilon_{c},  \\
    \epsilon_2 & =\epsilon_a+(1-\epsilon_a)\epsilon_b \approx \epsilon_{a}+\epsilon_{b}.
\end{split}
\end{equation}

All the multiplied terms, e.g. $\epsilon_{a}\epsilon_{b}$ in \eqref{eq.14} are ignored because the value is negligible.
Therefore, the successful offloading probability $P_{s}^{o}$ is given by
\begin{equation}\label{eq.15}
    P_{s}^{o}=(1-\epsilon_{1})(1-\epsilon_{2}).
\end{equation}

\subsection{Successful Execution Probability}\label{3.2}

As mentioned in Sec. \ref{2.1}, the transmission time for offloading is calculated as $t_{2}=NT_{s}$. When $t_{3}\leq T-t_{2}$, the successful execution probability is  $P_{s}^{e}=1$. Otherwise,  $P_{s}^{e}=0$ if $t_{3}\geq T-t_{2}$. Hence, we have $P^e_s=\mathds{1}(t_3 \leq T-t_2)$.
Therefore, the successful computation probability $P_{s}$ can be written as
\begin{equation}\label{eq.16}
    P_{s}= P_{s}^{o}P^e_s=\left\{
    \begin{aligned}
    & P_{s}^{o}, & t_{3}\leq T-t_{2}\\
    & 0, & t_{3}\geq T-t_{2}
    \end{aligned}
    \right.
\end{equation}

\textit{Remark 2}: If $\frac{M_{k}C_{\text{cpu}}}{f_{\text{user}}}\leq T$,  both users can finish their tasks locally within the delay budget so offloading is not needed and we have $P_{s}=1$.

\section{Problem Formulation and Algorithm}\label{4}

In this section, we aim to maximize the successful computation probability $P_{s}$ by optimizing the rate-splitting power allocation for splitting user, $P_{1,1}, ~P_{1,2}$, the offloading factor, $\lambda_{i}, i\in\{a,b,c\}$, and task-splitting factor, $\beta_{i}, i\in\{a,b,c\}$, after which offloading is performed. Thus, the problem is formulated as
\begin{subequations}\label{Prob.17}
    \begin{align}
        \max_{\textbf{m}, \bm{\lambda, \beta}
        } & \quad {P_{s}} \label{eq.17(a)}\\
        \mbox{s.t.} \quad
        & P_{1,1}+P_{1,2}\leq P_{t}, \label{eq.17(b)} \\
        & P_{2}\leq P_{t}, \label{eq.17(c)} \\
        & 0\leq\beta_{i}\leq 1, ~i\in\{a,b,c\} \label{eq.17(d)} \\      
        & t_{2}+t_{3}\leq T, \label{eq.17(e)}
    \end{align} 
\end{subequations}
where $\textbf{m}=[P_{1,1}, ~P_{1,2}, ~P_{2}]^{T}$ represents the transmit power of each stream, $\bm{\lambda}=[\lambda_{1}, ~\lambda_{2}]^{T}$ denotes the offloading factor of each user and $\bm{\beta}=[\beta_{a}, ~\beta_{b}, ~\beta_{c}]^{T}$ denotes task-splitting factor between the splitting streams. From \eqref{eq.15} and \eqref{eq.16}, we know that  
\begin{equation}\label{eq.18}
\begin{split}
    P_{s} & = (1-\epsilon_{1})(1-\epsilon_{2}) \\
    & \approx 1-\epsilon_{1}-\epsilon_{2} \\   
    & \approx 1-(2\epsilon_{a}+2\epsilon_{b}+\epsilon_{c}).
\end{split}
\end{equation}

Thus, Problem \eqref{Prob.17} is transformed into 
\begin{subequations}\label{Prob.19}
    \begin{align}
        \min_{\textbf{m}, \bm{\lambda, \beta}} & \quad {2\epsilon_{a}+2\epsilon_{b}+\epsilon_{c}} \label{eq.19(a)}\\
        \mbox{s.t.} \quad
        & \eqref{eq.17(b)}, \eqref{eq.17(c)}, \eqref{eq.17(d)}, \eqref{eq.17(e)}. \nonumber
    \end{align} 
\end{subequations}

According to the Chernoff bound of Q-function $\exp(-x^2/2)\geq Q(x), x\geq 0.5$, we have
\begin{equation}\label{eq.20}
    \exp{(\frac{-f^{2}(\gamma_{i}, M_{i})}{2})}\geq Q(f(\gamma_{i}, M_{i})), ~i\in\{a,b,c\}.   
\end{equation}

Therefore, we transform Problem \eqref{Prob.19} into Problem \eqref{Prob.21} to minimize the upper bound of the objective function.
\begin{subequations}\label{Prob.21}
    \begin{align}
        \min_{\textbf{m}, \bm{\lambda, \beta}} & \quad 2\exp{(\frac{-f^{2}(\gamma_{a}, M_{a})}{2})}+2\exp{(\frac{-f^{2}(\gamma_{b}, M_{b})}{2})} \nonumber \\
        & +\exp{(\frac{-f^{2}(\gamma_{c}, M_{c})}{2})}  \label{eq.21(a)}\\
        \mbox{s.t.} \quad
        & \eqref{eq.17(b)}, \eqref{eq.17(c)}, \eqref{eq.17(d)}, \eqref{eq.17(e)}.  \nonumber
    \end{align} 
\end{subequations}

Problem \eqref{Prob.21} is not convex and features strong coupling between variables in its objective function. For this kind of multi-variable-optimization problem, the classical \gls{ao} algorithms have been widely applied to decompose the original problem into separate sub-problems to update each variable iteratively until convergence. 

\subsection
{Offloading factor 
Optimization}

In this subsection, we optimize $\bm{\lambda}$ under fixed power allocation factor, $\textbf{m}$, and task-splitting factor, $\bm{\beta}$. The subproblem can be written as
\begin{subequations}\label{Prob.22}
    \begin{align}
        \min_{\bm{\lambda}} & \quad 2\exp{(\frac{-f^{2}(\gamma_{a}, M_{a})}{2})}+2\exp{(\frac{-f^{2}(\gamma_{b}, M_{b})}{2})} \nonumber\\
        & +\exp{(\frac{-f^{2}(\gamma_{c}, M_{c})}{2})} \label{eq.22(a)}\\
        \mbox{s.t.} \quad
        & t_{3}\leq T-t_{2}. \label{eq.22(b)}
    \end{align} 
\end{subequations}

According to \eqref{eq.7} and \eqref{eq.10}, constraint \eqref{eq.22(b)} can be rewritten as
\begin{subequations}
    \begin{align}
         & \frac{(1-\lambda_{k})M_{k}C_{\text{cpu}}}{f_{\text{user}}}\leq T-NT_{s}, ~k\in\{1,2\}\label{eq.23(a)} \\
         & \frac{\sum_{k=1}^{2}\lambda_{k}M_{k}C_{\text{cpu}}}{Lf_{\text{user}}}\leq T-NT_{s}, ~k\in\{1,2\}. \label{eq.23(b)}       
    \end{align}
\end{subequations}

Therefore, Problem \eqref{Prob.22} can be transformed into
\begin{subequations}\label{Prob.24}
    \begin{align}
        \min_{\bm{\lambda}} & \quad 2\exp{(\frac{-f^{2}(\gamma_{a}, M_{a})}{2})}+2\exp{(\frac{-f^{2}(\gamma_{b}, M_{b})}{2})} \nonumber\\
        & +\exp{(\frac{-f^{2}(\gamma_{c}, M_{c})}{2})} \label{eq.24(a)}\\
        & \eqref{eq.23(a)}, \eqref{eq.23(b)}. \nonumber
    \end{align} 
\end{subequations}

\textit{Lemma 1}: With fixed power allocation factor $\textbf{m}$ and task-splitting factor $\bm{\beta}$, the closed-form for $\lambda_{k}$ is given by 
\begin{equation}\label{eq.25}
    \lambda_{k}^{\star}=\max\{0,~1-\frac{(T-NT_{s})f_{\text{user}}}{M_{k}C_{\text{cpu}}}\}, ~k\in\mathcal{K}.
\end{equation}

\textit{Proof}: For arbitrary $\textbf{m}$ and $\bm{\beta}$, the objective function $2\exp{(\frac{-f^{2}(\gamma_{a}, M_{a})}{2})}+2\exp{(\frac{-f^{2}(\gamma_{b}, M_{b})}{2})}+\exp{(\frac{-f^{2}(\gamma_{c}, M_{c})}{2})}$ is clearly monotonically increasing \gls{wrt} $\exp{(\frac{-f^{2}(\gamma_{i}, M_{i})}{2})}, ~i\in\{a, b, c\}$. Besides, each exponential function in the objective function is monotonically decreasing \gls{wrt} $\lambda_{k}, k\in\mathcal{K}$. Therefore, the objective function in Problem \eqref{Prob.19} is monotonically increasing \gls{wrt} $\lambda_{k}$ (the proof of monotone is in Appendix 1). From constraint \eqref{eq.23(a)}, we can obtain the lower bound of $\lambda_{k}$
\begin{equation}\label{eq.26}
        1-\frac{(T-NT_{s})f_{\text{user}}}{M_{k}C_{\text{cpu}}} \leq \lambda_{k}, ~k\in\mathcal{K} .
\end{equation}

If $1-\frac{(T-NT_{s})f_{\text{user}}}{M_{k}C_{\text{CPU}}}< 0$, then we have $T-NT_{s}>\frac{M_{k}C_{\text{cpu}}}{f_{\text{user}}}$, which means each user can finish its own tasks within the time budget and it is not necessary to offload task to \gls{mec} server. As the objective function is monotonically increasing, the optimal value of $\lambda_{k}$ should be $\lambda_{k}^{\star}=\max\{0,~1-\frac{(T-NT_{s})f_{\text{user}}}{M_{k}C_{\text{cpu}}}\}, ~k\in\mathcal{K}$.

\subsection{Transmit Power 
Optimization}\label{4.2}
We now update $\mathbf{m}$ following an AO structure. For fixed  feasible task-splitting factor $\bm{\beta}$ and offloading factor $\bm{\lambda}$, we formulate the related subproblem by introducing slack variables $\textbf{t}=[t_{a}, ~t_{b}, ~t_{c}]^{T}$ and $\bm{\rho}=[\rho_{a}, ~\rho_{b}, ~\rho_{c}]^{T}$ as follows 
\begin{subequations}\label{Prob.27}
    \begin{align}
        \min_{\textbf{m},\textbf{t}, \bm{\rho}} & \quad 2\exp{(-t_{a})}+2\exp{(-t_{b})}+\exp{(-t_{c})}  \label{eq.27(a)}\\
        \mbox{s.t.} \quad
        & t_{i}\leq \frac{N[\log(1+\rho_{i})-\frac{M_{i}}{N}]^{2}}{2[1-(1+\rho_{i})^{-2}]}, ~i\in\{a,b,c\}\label{eq.27(b)} \\
        & \frac{P_{1,1}|h_{1}|^2} {P_{1,2}|h_{1}|^2+P_{2}|h_{2}|^2+\sigma_{n}^2}\geq \rho_{a}, \label{eq.27(c)} \\
        & \frac{P_{2}|h_{2}|^2}{P_{1,2}|h_{1}|^2+\sigma_{n}^2}\geq \rho_{b}, \label{eq.27(d)} \\
        & \frac{P_{1,2}|h_{1}|^2}{\sigma_{n}^2}\geq \rho_{c}, \label{eq.27(e)} \\
        &\eqref{eq.17(b)}, ~\eqref{eq.17(c)}. \nonumber
    \end{align} 
\end{subequations}

In the following, we address each complex non-convex constraint in Problem \eqref{Prob.27} sequentially. 

Constraint \eqref{eq.27(b)} is intractable, and we first rewrite it into 
\begin{equation}\label{eq.28}
    \frac{t_{i}}{N}(1-(1+\rho_{i})^{-2})\leq\frac{1}{2}\left(\log(1+\rho_{i})-\frac{M_{i}}{N}\right)^{2}.
\end{equation}

Since \eqref{eq.28} is not convex,
we first approximate the \gls{rhs} by its fist-order Taylor expression, which is given by 
\begin{equation}\label{eq.29}
    \frac{1}{2}\left(\log(1+\rho_{i})-\frac{M_{i}}{N}\right)^{2}\geq a_{i}^{[n]}\log(1+\rho_{i})+b_{i}^{[n]},
\end{equation}
where $a_{i}^{[n]}=\log(1+\rho_{i}^{[n]})-\frac{M_{i}}{N}$ and $b_{i}^{[n]}=\frac{1}{2}(a_{i}^{[n]})^{2}-a_{i}^{[n]}\log(1+\rho_{i}^{[n]})$. 

Similarly, we introduce an additional auxiliary variable, $\textbf{t}_{1}=[t_{1,a}, ~t_{1,b}, ~t_{1,c}]$, to approximate the \gls{lhs} of constraint \eqref{eq.28}. Firstly, we rewrite \gls{lhs} of \eqref{eq.28} as follows
\begin{equation}\label{eq.30}
    \frac{t_{i}}{N}(1-(1+\rho_{i})^{-2})=\frac{t_{i}}{N}-\frac{t_{i}(1+\rho_{i})^{-2}}{N},
\end{equation}
and we add the following constraints for the introduced variable:
\begin{equation}\label{eq.31}
    t_{1,i}\leq t_{i}(1+\rho_{i})^{-2}.
\end{equation}

Then \eqref{eq.31} is transformed into 
\begin{equation}\label{eq.32}
    \log(t_{1,i})+2\log(1+\rho_{i})\leq\log(t_{i}).
\end{equation}

Since \eqref{eq.32} is still not convex, $\log(t_{1,i})$ and $2\log(1+\rho_{i})$ are approximated by their first-order approximation around the point $\left(t_{1,i}^{[n]}\right)$ and $\left(\rho_{i}^{[n]}\right)$, which are given by
\begin{subequations}
    \begin{align}
         &\log(t_{1,i})  \leq\log(t_{1,i}^{[n]})+\frac{t_{1,i}-t_{1,i}^{[n]}}{t_{1,i}^{[n]}}, \label{eq.33(a)} \\
         &2\log(1+\rho_{i}) \leq 2\log(1+\rho_{i}^{[n]})+\frac{2(\rho_{i}-\rho_{i}^{[n]})}{1+\rho_{i}^{[n]}}. \label{eq.33(b)}
    \end{align}
\end{subequations}

Then the constraint \eqref{eq.28} can be rewritten into 
\begin{subequations}
    \begin{align}
        & \frac{t_{1}}{N}-\frac{t_{1,i}}{N}\leq a_{i}^{[n]}\log(1+\rho_{i})+b_{i}^{[n]}, \label{eq.34(a)}\\
        & \log(t_{1,i}^{[n]})+\frac{t_{1,i}-t_{1,i}^{[n]}}{t_{1,i}^{[n]}}+2\log(1+\rho_{i}^{[n]})+\frac{2(\rho_{i}-\rho_{i}^{[n]})}{1+\rho_{i}^{[n]}} \nonumber \\
        & \quad\quad\leq\log(t_{i}
        )\label{eq.34(b)}.
    \end{align}
\end{subequations}

To handle the non-convex \eqref{eq.27(c)} and \eqref{eq.27(d)}, we rewrite them in their Difference-of-Convex forms, which are given by 
\begin{subequations}
    \begin{align}
        & P_{1,2}|h_{1}|^2+P_{2}|h_{2}|^2+\sigma_{n}^2-\frac{P_{1,1}|h_{1}|^2}{\rho_{a}}\leq 0, \label{eq.35(a)} \\
        & P_{1,2}|h_{1}|^2+\sigma_{n}^2-\frac{P_{2}|h_{2}|^2}{\rho_{b}}\leq 0. \label{eq.35(b)}
    \end{align}
\end{subequations}

\eqref{eq.35(a)} and \eqref{eq.35(b)} still have concave parts, i.e., $-\frac{P_{1,1}|h_{1}|^2}{\rho_{a}}$ and $-\frac{P_{2}|h_{2}|^2}{\rho_{b}}$, which are rewritten by the first-order Taylor approximations. Specifically, the constraints \eqref{eq.35(a)} and \eqref{eq.35(b)} are respectively approximated around the point ($\mathbf{m}^{[n]}, \bm{\rho}^{[n]}$) at iteration $n$ by 
\begin{subequations}
    \begin{align}
        & P_{1,2}|h_{1}|^2+P_{2}|h_{2}|^2+\sigma_{n}^2 \nonumber \\
        & \quad\quad-\frac{P_{1,1}|h_{1}|^2}{\rho_{a}^{[n]}}+(\rho_{a}-\rho_{a}^{[n]})\frac{P_{1,1}^{[n]}|h_{1}|^2}{(\rho_{a}^{[n]})^2} \leq 0, \label{eq.36(a)} \\
        & P_{1,2}|h_{1}|^2+\sigma_{n}^2 -\frac{P_{2}|h_{2}|^2}{\rho_{b}^{[n]}}+(\rho_{b}-\rho_{b}^{[n]})\frac{P_{2}^{[n]}|h_{2}|^2}{(\rho_{b}^{[n]})^2} \leq 0. \label{eq.36(b)}       
    \end{align}
\end{subequations}

The non-convex subproblem \eqref{Prob.27} is transformed into a convex Problem \eqref{Prob.37} based on the aforementioned approximation techniques, which can then be solved via the \gls{sca} approach. By approximating a series of convex sub-problems, \gls{sca} solves the original problem. At iteration $n$, we solve the following problem using the best solution $(\textbf{m}^{[n-1]},\bm{\rho}^{[n-1]})$ from the preceding iteration $n-1$:
\begin{subequations}\label{Prob.37}
    \begin{align}
        \min_{\textbf{m}, \bm{\rho}} & \quad 2\exp{(-t_{a})}+2\exp{(-t_{b})}+\exp{(-t_{c})}  \label{eq.37(a)}\\
        \mbox{s.t.} \quad
        & \eqref{eq.34(a)}, \eqref{eq.34(b)}, \eqref{eq.35(a)},
        \eqref{eq.35(b)}
        \eqref{eq.27(e)}
        \eqref{eq.17(b)}, \eqref{eq.17(c)}. \nonumber
    \end{align} 
\end{subequations}

\subsection{Rate-Splitting Task Allocation Factor 
Optimization}\label{4.3}
When transmit power $\textbf{m}$ and the offloading factor $\bm{\lambda}$ are all fixed, the subproblem with respect to $\bm{\beta
}$ is given by

\begin{subequations}\label{Prob.38}
    \begin{align}
        \max_{\bm{\beta}} & \quad 2\exp{(-t_{a})}+2\exp{(-t_{b})} +\exp{(-t_{c})}  \label{eq.38(a)}\\
        \mbox{s.t.} \quad
        & t_{i}\leq \frac{N[\log(1+\rho_{i})-\frac{\beta_{i}\lambda_{k}M_{k}}{N}]^{2}}{2[1-(1+\rho_{i})^{-2}]}, ~i\in\{a,b,c \}, ~k\in\mathcal{K} \label{eq.38(b)} \\
        & \eqref{eq.17(d)}. \nonumber
    \end{align} 
\end{subequations}

\textit{Remark 3}: Notice that in \eqref{eq.38(b)}, the subscription $i=\{a,c\}$ corresponds to $k=1$ and $i=b$ corresponds to $k=2$ based on the definition of $M_{i}$ in Sec. \ref{3.1}. We rewrite constraint \eqref{eq.38(b)} as
\begin{equation} \label{eq.39}
    t_{i}\leq Nc_{i}(d_{i}+e_{i}\beta_{i})^{2},
\end{equation}
where $c_{i}\triangleq \frac{1}{2[1-(1+\rho_{i})^{-2}]}$ and $d_{i}\triangleq\log(1+\rho_{i})$ and $e_{i}\triangleq -\frac{\lambda_{i}M_{k}}{N}$.

Similar to \eqref{eq.27(b)}, constraint \eqref{eq.39} is not convex, $(e_{i}\beta_{i})^2$ is approximated by its first-order Taylor approximation around the point $\left(\bm{\beta}^{[n]}\right)$ at iteration $n$ which is given by
\begin{equation}\label{eq.40}
    (e_{i}\beta_{i})^2 \geq (e_{i}\beta_{i}^{[n]})^2+2(\beta_{i}-\beta_{i}^{[n]})(e_{i}^{2}\beta_{i}^{[n]})\triangleq\Phi(\beta_{i}^{[n]}).
\end{equation}

Thus, constraint \eqref{eq.38(b)} is rewritten into 
\begin{equation}\label{eq.41}
    t_{i}\leq Nc_{i}\left(d_{i}^{2}+2d_{i}e_{i}\beta_{i}+\Phi(\beta_{i}^{[n]})\right).
\end{equation}

Therefore, subproblem \eqref{Prob.38} is transformed into 
\begin{subequations}\label{Prob.42}
    \begin{align}
        \min_{\bm{\beta}} & \quad 2\exp{(-t_{a})}+2\exp{(-t_{b})} +\exp{(-t_{c})} \label{eq.42(a)}\\
        \mbox{s.t.} \quad
        & \eqref{eq.41}, \eqref{eq.17(d)}.
    \end{align} 
\end{subequations}

Now, it is easy to verify that all these sub-problems Problem \eqref{Prob.37} and Problem \eqref{Prob.42} are convex problems, which can be efficiently solved by standard convex problem solver such as CVX \cite{grant2009cvx}.

\subsection{Proposed Algorithm, Convergence and Complexity}
According to the above three subproblems, \gls{ao} is applied to solve Problem \eqref{Prob.21} by utilizing the \gls{sca}-based algorithm. Specifically, the offloading factor $\lambda$, transmit power $\textbf{m}$ and the task-splitting factor $\bm{\beta}$ are alternately optimized by solving Problem \eqref{Prob.37} and \eqref{Prob.42}, respectively. The details of the proposed algorithm are summarised in \textbf{Algorithm 1} where $\tau$ is the tolerance. Define $\epsilon=2\exp{(-t_{a})}+2\exp{(-t_{b})}+\exp{(-t_{c})}$ for convenience, and $\epsilon^{[n]}=2\exp{(-t_{a}^{[n]})}+2\exp{(-t_{b}^{[n]})}+\exp{(-t_{c}^{[n]})}$. At each iteration of Algorithm 1, the power allocation and task-splitting factor are updated by solving Second Order Cone Programming (SOCP) problems. Each SOCP is solved by using interior-point method with the computational complexity of $\mathcal{O}\left([X]^{3.5}\right)$, where $X$ is the total number of variables in the corresponding SOCP problem. Although an additional variable $\bm{\rho}$, $\bm{t}$ and $\bm{t}_{1}$ are introduced for approximating convex problems, the main complexity still comes from the power allocation and task-splitting factor optimization. With given task-splitting factor, the number of variables of Problem \eqref{Prob.37} is given by $X_{\text{power allocation}}=\left(K+1\right)$. Similarly, the number of variables of Problem \eqref{Prob.42} is given by $X_{\text{task-splitting facotr}}=\left(K+1\right)$. The total number of iterations required for the convergence is $\mathcal{O}\left(\log(\tau^{-1})\right)$, where $\tau$ is the convergence tolerance of Algorithm 1. Therefore, the total computation complexity of Algorithm 1 is $\mathcal{O}\left((K+1)^{3.5}\log(\tau^{-1})\right)$.

\begin{algorithm} [t!] \label{Alg.1}
\caption{Proposed \gls{sca}-based AO algorithm for solving Problem \eqref{Prob.21}}
\LinesNumbered
\SetKwInput{kwInit}{Initialise}
\kwInit{$n\gets0, ~\epsilon^{[n]}\gets 0, \text{and feasible} ~\textbf{m}^{[n]}, ~\bm{\beta}^{[n]}$;\\
\text{Calculate} optimal $\lambda_{k}^{*}, k\in\{1, 2\}$ \text{by \textit{Lemma 1}.}}
\Repeat{$|\epsilon^{[n]}-\epsilon^{[n-1]}|\leq \tau$}
 {$n\gets n+1$; \\
 Find optimal $\textbf{m}^{[n]}$ by solving Problem \eqref{Prob.37} for given $\bm{\lambda}^{[n]}$ and $\bm{\beta}^{[n-1]}$; \\
 Find optimal $\bm{\beta}^{[n]}$ by solving Problem \eqref{Prob.42} for given $\bm{\lambda}^{[n]}$ and $\textbf{m}^{[n]}$; \\
 Update $\epsilon^{[n]}\gets \epsilon^{*}, ~\bm{\lambda}^{[n]}\gets \bm{\lambda}^{*}, ~\textbf{m}^{[n]}\gets\textbf{m}^{*}, ~\bm{\beta}^{[n]}\gets\bm{\beta}^{*}$;}
\end{algorithm}

Next we demonstrate the convergence of Algorithm 1. Define $g\left(\textbf{m}^{[n]}, \bm{\beta}^{[n]}\right)$ as the objective value at the $n^{\text{th}}$ iteration. First, from Problem \eqref{Prob.37} with a given $\bm{\lambda}$ and $\bm{\beta}$ in step 3 of Algorithm 1, we know
\begin{equation} \label{eq.43}
    \begin{split}
        g\left(\textbf{m}^{[n-1]}, \bm{\beta}^{[n-1]}\right) & \overset{a}{=} g_{\textbf{m}}\left(\textbf{m}^{[n-1]}, \bm{\beta}^{[n-1]}\right) \\
        & \overset{b}{\leq} g_{\textbf{m}}\left(\textbf{m}^{[n]}, \bm{\beta}^{[n-1]}\right) \\
        & \overset{c}{\leq} g\left(\textbf{m}^{[n]}, \bm{\beta}^{[n-1]}\right),
    \end{split}
\end{equation}
where $g_{\textbf{m}}$ represents the objective value of Problem \eqref{Prob.37}. $\textit{a}$ holds since the first-order Taylor approximations are tight at the given point $\left(\textbf{m}^{[n-1]}, \bm{\beta}^{[n-1]}\right)$. Since the solution of the approximated Problem \eqref{Prob.37} at the $n-1^{\text{th}}$ iteration is a feasible point for Problem \eqref{Prob.37} at the $n^{\text{th}}$ iteration, it can be solved successfully. Moreover, the objective function is bounded by the transmit power constraints, then \textit{b} holds. \textit{c} is due to the objective value of \eqref{Prob.37} being the lower bound of \eqref{Prob.27}.

Similarly, from Problem \eqref{Prob.42} with a given $\bm{\lambda}$ and $\bm{m}$ in step 4 of Algorithm 1, we know
\begin{equation} \label{eq.44}
    \begin{split}
        g\left(\textbf{m}^{[n]}, \bm{\beta}^{[n-1]}\right) & \overset{a}{=} g_{\bm{\beta}}\left(\textbf{m}^{[n]}, \bm{\beta}^{[n-1]}\right) \\
        & \overset{b}{\leq} g_{\bm{\beta}}\left(\textbf{m}^{[n]}, \bm{\beta}^{[n]}\right) \\
        & \overset{c}{\leq} g\left(\textbf{m}^{[n]}, \bm{\beta}^{[n]}\right),
    \end{split}
\end{equation}
where $g_{\bm{\beta}}$ represents the objective value of Problem \eqref{Prob.42}. $\textit{a}$, $\textit{b}$ and $\textit{c}$ hold for the same reasons as we described previously.
Thus, we can obtain that
\begin{equation} \label{eq.45}
    g\left(\textbf{m}^{[n-1]}, \bm{\beta}^{[n-1]}\right)\leq g\left(\textbf{m}^{[n]}, \bm{\beta}^{[n]}\right),
\end{equation}
which proves that Algorithm 1 generates a non-decreasing sequence of objective values and it is bounded by the power budget.

\textit{Remark 4}: The proposed system model and algorithm are designed for a \gls{siso} two-user system but can be extended to a general \gls{mimo} $K$-user system. In this extension, the expression for \gls{scp} must be revised. Specifically, from \eqref{eq.15} and \eqref{eq.16}, the \gls{scp} in the two-user system is given as $P_{s} = \prod_{k=1}^{2}(1-\epsilon_{k})$. By generalizing to a $K$-user system, the \gls{scp} expression becomes $P_{s} = \prod_{k=1}^{K}(1-\epsilon_{k})$. In addition, each $\epsilon_{k}$ needs to be re-derived, accounting for the effects of error propagation in a similar way. If extended to multiple-antenna scenarios, precoder design shall be considered at the transmitter side instead of power allocation. The algorithm for an uplink \gls{mimo} \gls{rsma} system in the \gls{fbl} regime can be referred to \cite{10741240}.

\section{Numerical Results}\label{5}
The performance of a \gls{fbl} \gls{rsma}-aided \gls{mec} system with two users is evaluated and compared with conventional transmission schemes in this section. The \gls{scp} performance of \gls{rsma}-aided \gls{mec} is illustrated. Here, the performance of \gls{rsma} is compared to \gls{noma}. 
\begin{itemize}  
\item \gls{noma}: This is a special case of \gls{rsma} where none of the users split their messages in \gls{noma}. The BS sequentially decodes the user messages based on \gls{sic}. 
\end{itemize}
Simulations for the Rayleigh Fading channel with 100 channel realisations are performed. The time budget, $T$, is 10 ms and the time length of blocklength (in symbol), $T_s$, is 0.0025 ms. The CPU computation is $C_{\text{cpu}} = 1000$ cycles/bits, $f_{\text{user}}= 0.5$ GHz and $L$ is assumed to be 5.  Without loss of generality, it is assumed that the noise variance is 1, $\sigma_{n}^2=1$. The tolerance of the algorithm is set to be $\tau=10^{-3}$. Here, the decoding orders of $s_{1,1}\to s_2\to s_{1,2}$ of \gls{rsma} and $s_1\to s_2$ of \gls{noma} are chosen, respectively according to \cite{9970313}. 

\subsection{The SCP versus Task Size}\label{5.1}
\begin{figure}[t!]
    \centering
    \includegraphics[scale=0.65]{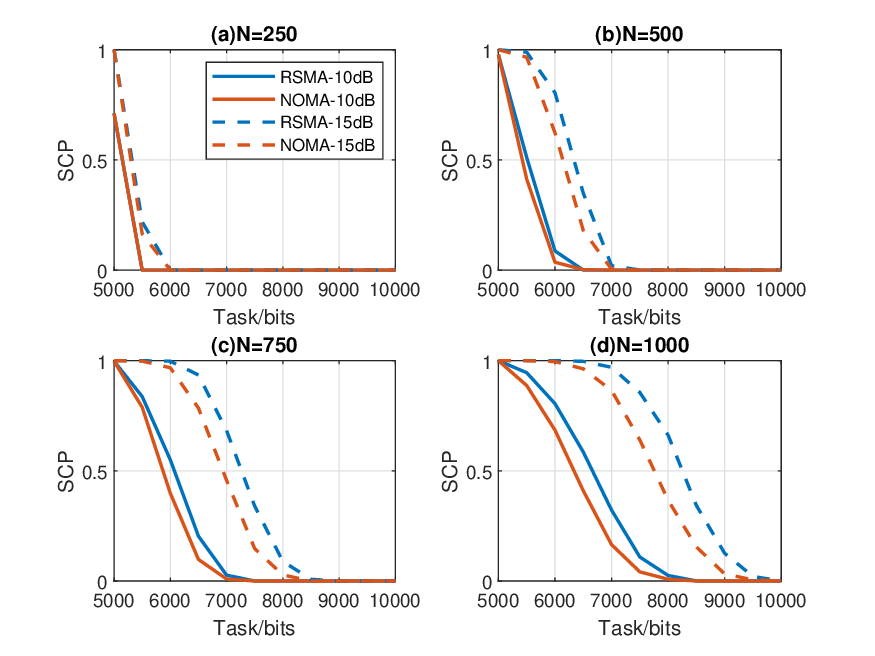}
    \caption{The successful computation probability performance of RSMA and NOMA versus the size of tasks with two different transmit SNR averaged over 100 random channel realisations. $M_{2}=5.5$k bits. (a) N=250; (b) N=500; (c) N=750; (c) N=1000.}
    \label{Fig.2}
\end{figure}

The trend of \gls{scp} as the size of tasks increases is shown in Fig. \ref{Fig.2}. Results of four different blocklengths, $N=250, 500, 750$ and $1000$, are compared. In this simulation, the task size of user-1, $M_{1}$, is from 5k bits to 10k bits and the task size of user-2 is $M_{2}=5.5$k bits. Solid and dash lines represent the transmit \gls{snr} of 10 dB and 15 dB, respectively. Recall that when $\frac{M_{k}C_{\text{cpu}}}{f_{\text{user}}} \leq T, ~k \in {1, 2}$, user-$k$ can complete its tasks locally within the time budget, making offloading unnecessary. For user-$1$, with $M_{1} = 5$k bits, the local computation time is $\frac{M_{1}C_{\text{cpu}}}{f_{\text{user}}} = 10$ ms $\leq T$, so offloading is not required. In contrast, user-$2$ has a local computation time of $\frac{M_{2}C_{\text{cpu}}}{f_{\text{user}}} = 11$ ms $> T$, necessitating offloading to meet the time budget. Consequently, transmission errors occur only during user-$2$'s offloading process when $M_{1} = 5$k bits and $M_{2} = 5.5$k bits. As $M_{1}$ increases, longer local computation time is required for user-$1$, and offloading to the \gls{mec} server becomes necessary for user-$1$. We now summarize the insights from Fig. \ref{Fig.2} as follows. 

\begin{itemize}
\item Fig. \ref{Fig.2} firstly illustrates the enhanced SCP performance as SNR increases, comparing the solid line (SNR=10 dB) and the dashed line (SNR=15 dB). For example, in Fig. \ref{Fig.2} (a), with a short blocklength of 250,  the \gls{scp} of both \gls{rsma} and \gls{noma} drops to 0 when the task size reaches and goes beyond 5.5k bits given a transmit \gls{snr} of 10 dB. This is because the short blocklength results in a significantly high coding rate in FBL, which, however, can be mitigated by an increased SNR. As a verification, in Fig. \ref{Fig.2} (a), when the transmit \gls{snr} increases to 15 dB, the \gls{scp} is approximately 0.2 (not zero as a comparison) for $M_{1} = 5.5$k bits and drops to 0 when $M_{1}= 6$k bits. 

\item Secondly, Fig. \ref{Fig.2} demonstrates the superiority of  \gls{rsma} over \gls{noma} in the \gls{mec} scheme. For example, Fig. \ref{Fig.2} (b), with a task size of 6k bits and a transmit \gls{snr} of 15 dB, \gls{rsma} achieves a significantly higher \gls{scp} of approximately 0.8 with a blocklength of 500, as shown in Fig. \ref{Fig.2}, compared to \gls{noma}, which achieves an \gls{scp} of about 0.6. At a \gls{scp} of 0.5 and a blocklength of 750, the maximum task sizes for \gls{rsma} are approximately 6.1k bits and 7.4k bits, compared to 5.8k bits and 7k bits for \gls{noma}, at transmit SNRs of 10 dB and 15 dB, respectively, as shown in Fig. \ref{Fig.2} (c).

The superiority of \gls{rsma} over \gls{noma} comes from the higher flexibility in decoding order and in allocating power to each stream. The streams of the second decoded user would have a considerable impact on the first decoded user's \gls{sinr} since the decoding order strictly limits \gls{noma}. However, the decoding order can be more flexible to balance the \gls{sinr} while decoding each stream with the help of \gls{rsma} by splitting one user, for instance, the decoding order for \gls{rsma} is $s_{1,1} \to s_{2} \to s_{1,2}$. In this approach, the task $M_{1}$ for user-$1$ is split into two parts which are transmitted by $s_{1,1}$ and $s_{1,2}$. The majority of $M_{1}$, along with a significant portion of user-$1$'s transmit power, is allocated to $s_{1,2}$, which is decoded interference-free after $s_{2}$. This strategy significantly reduces the decoding burden on $s_{1,1}$, enabling more efficient interference management compared to \gls{noma}, where the entire task $M_{1}$ must contend with interference. By leveraging task splitting and optimizing the power allocation, \gls{rsma} achieves better resource utilization and improves the overall system performance.

\item Finally,  Fig. \ref{Fig.2} (a) - Fig. \ref{Fig.2} (d) demonstrates an enlarged performance gain of \gls{rsma} over \gls{noma} with the increase of blocklength. Especially in Fig. \ref{Fig.2} (d), for \gls{rsma}, the maximum task size that user-$1$ can handle is 6.7k bits at a transmit SNR of 10 dB and 8.2k bits at 15 dB. In comparison, for \gls{noma}, the maximum task size is 6.5k bits at 10 dB and 7.8k bits at 15 dB. 

Additionally, the offloading time, $t_{2} = NT_{s}$, increases with the blocklength, reducing the available computation time, $t_{3}$. As a result, the user must offload a larger portion of tasks to the \gls{mec} server, leading to a higher coding rate. In this scenario, \gls{rsma} proves advantageous by splitting a user’s tasks, effectively balancing the \gls{sinr} of each stream and reducing the offloading error probability.

\end{itemize}

\subsection{The SCP versus Blocklength}\label{5.2}
\begin{figure}[t!]
    \centering
    \includegraphics[scale=0.65]{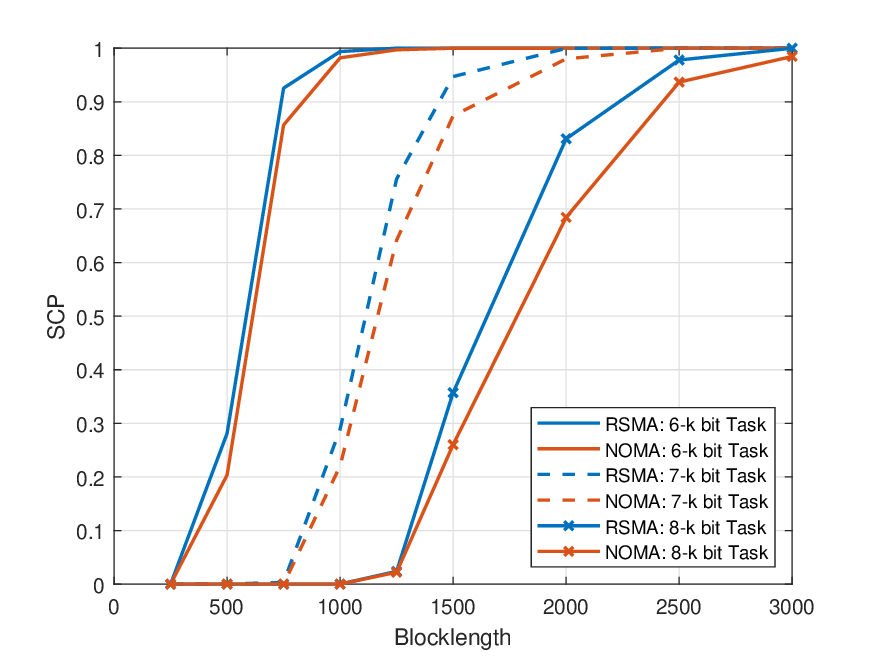}
    \caption{The successful computation probability performance of RSMA and NOMA versus blocklength with two different task sizes averaged over 100 random channel realisations. Transmit SNR is 15 dB.}
    \label{Fig.3}
\end{figure}

Fig. \ref{Fig.3} illustrates the relationship between \gls{scp} and blocklength. The comparison is made for two task sizes, $M_{k} = 6$k bits, $M_{k} = 7$k bits and $M_{k} = 8$k bits, where $k \in \{1, 2\}$. The transmit SNR is fixed at 15 dB for this simulation. \gls{rsma} requires a shorter blocklength than \gls{noma} to achieve the same \gls{scp}. For example, when the task size for both users is 6k bits, \gls{rsma} achieves an \gls{scp} of 0.5 with 50 fewer blocklengths  (in symbol) compared to \gls{noma} as indicated by the two solid lines. This demonstrates that \gls{rsma} can effectively reduce latency by requiring less blocklength. As the task size increases, the \gls{scp} performance gain of \gls{rsma} over \gls{noma} becomes more significant, as illustrated by the other two groups of lines. For larger task sizes, the coding rate increases proportionally, placing a heavy burden on decoding $s_{1}$ at the \gls{mec} server, as its \gls{sinr} is heavily influenced by $s_{2}$ for \gls{noma}. To decode $s_{2}$ at a high coding rate, more power is allocated to $s_{2}$, which results in an increasing interference experienced by $s_{1}$. 

\subsection{The SCP versus Transmit SNR}\label{5.3}
\begin{figure}[t!]
    \centering
    \includegraphics[scale=0.65]{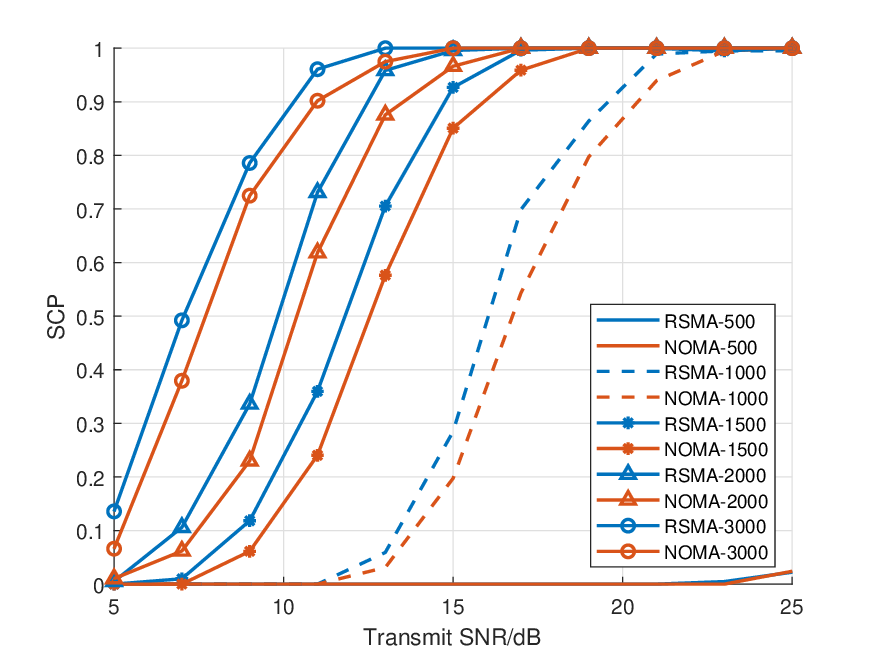}
    \caption{The successful computation probability performance of RSMA and NOMA versus transmit SNR with four different finite blocklength averaged over 100 random channel realisations. $M_{1}=7$k bits. $M_{2}=7$k bits.}
    \label{Fig.4}
\end{figure}

Fig. \ref{Fig.4} shows the trend of \gls{scp} as the transmit \gls{snr} increases with more comprehensive simulations. The comparison includes results for five specific blocklengths: $N = 500$, $1000$, $1500$, $2000$ and $3000$. In this simulation, both user-$1$ ($M_{1}$) and user-$2$ ($M_{2}$) have task sizes of 7k bits. The blue lines represent \gls{rsma}, while the red lines correspond to \gls{noma}.

In Fig. \ref{Fig.4}, with blocklength of 500, neither \gls{rsma} and \gls{noma} can successfully complete own tasks within the time budget (\gls{scp} is 0) since the coding rate is too high. As the blocklength increases, the \gls{scp} performance improves with the \gls{scp} of \gls{noma} remains consistently lower than \gls{rsma}. The gain of \gls{rsma} over \gls{noma} does not continue to increase with blocklength, instead, it saturates at the blocklength of approximately 1000, revealing the performance limit of \gls{rsma}. 

These numerical results demonstrates that \gls{rsma} with a decoding order of $s_{1,1} \to s_{2} \to s_{1,2}$ achieves a higher \gls{scp} compared to \gls{noma} with a decoding order of $s_{1} \to s_{2}$ in \gls{fbl} regime. The superior performance of \gls{rsma} can be attributed to its ability to allocate transmit power and offloading task size between split streams, balancing the \gls{sinr} and enhancing overall \gls{scp} performance.

\section{Conclusion}\label{6}
This paper introduces an \gls{rsma}-aided \gls{mec} system operating in the \gls{fbl} regime, where \gls{scp} serves as a critical metric. The study focuses on a two-user \gls{rsma}-aided \gls{mec} system, aiming to optimize power allocation, task-splitting factor, and offloading factors to maximize \gls{scp}. Emphasizing the scenario where only one user's task is split, simulations demonstrate that the \gls{rsma}-aided \gls{mec} system achieves significantly higher \gls{scp} compared to traditional schemes such as \gls{noma} with \gls{fbl} constraints. Furthermore, \gls{rsma} enables reduced latency by achieving comparable \gls{scp} to \gls{noma} with shorter blocklengths. These findings shows \gls{rsma}-assisted \gls{mec} as a promising approach for reliable, low-latency wireless communication systems in the future.

In conclusion, \gls{rsma} continuously produces a higher \gls{scp} than \gls{noma} in \gls{fbl} regime, allowing \gls{rsma} to provide a more dependable performance. Besides, \gls{rsma} achieves much lower latency while achieving the same \gls{scp} as \gls{noma} with a shorter blocklength. Since this work is based on a two-user \gls{siso} system, the general $K$-user \gls{mimo} system can be studied in the future.

\appendix
It is easy to note that the objective function $2\exp{(\frac{-f^{2}(\gamma_{a}, M_{a})}{2})}+2\exp{(\frac{-f^{2}(\gamma_{b}, M_{b})}{2})}+\exp{(\frac{-f^{2}(\gamma_{c}, M_{c})}{2})}$ is clearly monotonically increasing with respect to (w.r.t) $\exp{(\frac{-f^{2}(\gamma_{i}, M_{i})}{2})}, ~i\in\{a, b, c\}$, for arbitrary $\textbf{m}$ and $\bm{\beta}$. To verify the monotone of $\exp{(\frac{-f^{2}(\gamma_{i}, M_{i})}{2})}, ~i\in\{a, b, c\}$, we first find the derivative of $f(\gamma_{i}, M_{i})^2$. We take $f(\gamma_{1}, M_{a})^2$ for example. First we rewrite $f(\gamma_{1}, M_{a})$ as
\begin{equation}
    f(\gamma_{1}, M_{a})=\frac{\log(1+\gamma_{1})-\frac{\lambda_{1}M_{1}}{N}}{D},
\end{equation}
where $D=\sqrt{(1-(1+\gamma)^{-2})/N}$. Then the first-order derivative of $f(\gamma_{1}, M_{a})^2$ is
\begin{equation}
    \frac{\partial f(\gamma_{1}, M_{a})^2}{\partial \lambda_{1}}=-\frac{2M_{1}}{DN}(\log(1+\gamma_{1})-\frac{\lambda_{1}M_{1}}{N})\leq0,
\end{equation}
since we have $\log(1+\gamma_{1})-\frac{\lambda_{1}M_{1}}{N}\geq 0$. As $\exp{(\frac{-f^{2}(\gamma_{1}, M_{a})}{2})}$ is monotonically decreasing with $f(\gamma_{1}, M_{a})^2$ and $f(\gamma_{1}, M_{a})^2$ is monotonically decreasing \gls{wrt} $\lambda_{1}$, then $\exp{(\frac{-f^{2}(\gamma_{1}, M_{a})}{2})}$ is monotonically increasing \gls{wrt} $\lambda_{1}$. The proofs are similar for $\exp{(\frac{-f^{2}(\gamma_{2}, M_{b})}{2})}$ and $\exp{(\frac{-f^{2}(\gamma_{3}, M_{a})}{c})}$. Therefore, the objective function is also monotonically increasing \gls{wrt} each $\lambda$.

\bibliographystyle{ieeetr}
\bibliography{ref}

\begin{thebibliography}{10}

\bibitem{Li2019}
R.~Li, ``{Network 2030 A Blueprint of Technology, Applications and Market Drivers Towards the Year 2030 and Beyond},'' tech. rep., International Telecommunication Union (ITU), 2019.

\bibitem{8766143}
Z.~Zhang, Y.~Xiao, Z.~Ma, M.~Xiao, Z.~Ding, X.~Lei, G.~K. Karagiannidis, and P.~Fan, ``{6G Wireless Networks: Vision, Requirements, Architecture, and Key Technologies},'' {\em IEEE Vehicular Technology Magazine}, vol.~14, no.~3, pp.~28--41, 2019.

\bibitem{8016573}
Y.~Mao, C.~You, J.~Zhang, K.~Huang, and K.~B. Letaief, ``{A Survey on Mobile Edge Computing: The Communication Perspective},'' {\em IEEE Communications Surveys \& Tutorials}, vol.~19, no.~4, pp.~2322--2358, 2017.

\bibitem{7879258}
P.~Mach and Z.~Becvar, ``{Mobile Edge Computing: A Survey on Architecture and Computation Offloading},'' {\em IEEE Communications Surveys \& Tutorials}, vol.~19, no.~3, pp.~1628--1656, 2017.

\bibitem{abbas2017mobile}
N.~Abbas, Y.~Zhang, A.~Taherkordi, and T.~Skeie, ``{Mobile Edge Computing: A Survey},'' {\em IEEE Internet of Things Journal}, vol.~5, no.~1, pp.~450--465, 2017.

\bibitem{etsi}
ETSI, ``{Mobile-Edge Computing – Introductory Technical White Paper}.'' \url{https://portal.etsi.org/Portals/0/TBpages/MEC/Docs/Mobile-edge_Computing_-_Introductory_Technical_White_Paper_V1%2018-09-14.pdf}, Sep. 2014.
\newblock Accessed at 15/12/2024.

\bibitem{you2016energy}
C.~You, K.~Huang, H.~Chae, and B.-H. Kim, ``{Energy-Efficient Resource Allocation for Mobile-Edge Computation Offloading},'' {\em IEEE Transactions on Wireless Communications}, vol.~16, no.~3, pp.~1397--1411, 2016.

\bibitem{li2021energy}
X.~Li, R.~Fan, H.~Hu, N.~Zhang, X.~Chen, and A.~Meng, ``{Energy-Efficient Resource Allocation for Mobile Edge Computing with Multiple Relays},'' {\em IEEE Internet of Things Journal}, vol.~9, no.~13, pp.~10732--10750, 2021.

\bibitem{sun2019joint}
H.~Sun, F.~Zhou, and R.~Q. Hu, ``{Joint Offloading and Computation Energy Efficiency Maximization in A Mobile Edge Computing System},'' {\em IEEE Transactions on Vehicular Technology}, vol.~68, no.~3, pp.~3052--3056, 2019.

\bibitem{saito2013non}
Y.~Saito, Y.~Kishiyama, A.~Benjebbour, T.~Nakamura, A.~Li, and K.~Higuchi, ``{Non-Orthogonal Multiple Access (NOMA) for Cellular Future Radio Access},'' in {\em 2013 IEEE 77th vehicular technology conference (VTC Spring)}, pp.~1--5, IEEE, 2013.

\bibitem{islam2016power}
S.~R. Islam, N.~Avazov, O.~A. Dobre, and K.-S. Kwak, ``{Power-Domain Non-Orthogonal Multiple Access (NOMA) in 5G Systems: Potentials and Challenges},'' {\em IEEE Communications Surveys \& Tutorials}, vol.~19, no.~2, pp.~721--742, 2016.

\bibitem{khan2019joint}
W.~U. Khan, F.~Jameel, T.~Ristaniemi, S.~Khan, G.~A.~S. Sidhu, and J.~Liu, ``{Joint Spectral and Energy Efficiency Optimization for Downlink NOMA Networks},'' {\em IEEE Transactions on Cognitive Communications and Networking}, vol.~6, no.~2, pp.~645--656, 2019.

\bibitem{khan2020spectral}
W.~U. Khan, J.~Liu, F.~Jameel, V.~Sharma, R.~J{\"a}ntti, and Z.~Han, ``{Spectral Efficiency Optimization for Next Generation NOMA-Enabled IoT Networks},'' {\em IEEE Transactions on Vehicular Technology}, vol.~69, no.~12, pp.~15284--15297, 2020.

\bibitem{8537962}
F.~Wang, J.~Xu, and Z.~Ding, ``{Multi-Antenna NOMA for Computation Offloading in Multiuser Mobile Edge Computing Systems},'' {\em IEEE Transactions on Communications}, vol.~67, no.~3, pp.~2450--2463, 2019.

\bibitem{8673584}
Z.~Ding, J.~Xu, O.~A. Dobre, and H.~V. Poor, ``{Joint Power and Time Allocation for NOMA–MEC Offloading},'' {\em IEEE Transactions on Vehicular Technology}, vol.~68, no.~6, pp.~6207--6211, 2019.

\bibitem{zhu2020resource}
J.~Zhu, J.~Wang, Y.~Huang, F.~Fang, K.~Navaie, and Z.~Ding, ``{Resource Allocation for Hybrid NOMA MEC Offloading},'' {\em IEEE Transactions on Wireless Communications}, vol.~19, no.~7, pp.~4964--4977, 2020.

\bibitem{9179779}
F.~Fang, Y.~Xu, Z.~Ding, C.~Shen, M.~Peng, and G.~K. Karagiannidis, ``{Optimal Resource Allocation for Delay Minimization in NOMA-MEC Networks},'' {\em IEEE Transactions on Communications}, vol.~68, no.~12, pp.~7867--7881, 2020.

\bibitem{8492422}
Z.~Ding, D.~W.~K. Ng, R.~Schober, and H.~V. Poor, ``{Delay Minimization for NOMA-MEC Offloading},'' {\em IEEE Signal Processing Letters}, vol.~25, no.~12, pp.~1875--1879, 2018.

\bibitem{ye2020enhance}
Y.~Ye, R.~Q. Hu, G.~Lu, and L.~Shi, ``{Enhance Latency-Constrained Computation in MEC Networks Using Uplink NOMA},'' {\em IEEE Transactions on Communications}, vol.~68, no.~4, pp.~2409--2425, 2020.

\bibitem{9905567}
Z.~Liu, Y.~Zhu, Y.~Hu, P.~Sun, and A.~Schmeink, ``{Reliability-Oriented Design Framework in NOMA-Assisted Mobile Edge Computing},'' {\em IEEE Access}, vol.~10, pp.~103598--103609, 2022.

\bibitem{9831440}
Y.~Mao, O.~Dizdar, B.~Clerckx, R.~Schober, P.~Popovski, and H.~V. Poor, ``{Rate-Splitting Multiple Access: Fundamentals, Survey, and Future Research Trends},'' {\em IEEE Communications Surveys \& Tutorials}, vol.~24, no.~4, pp.~2073--2126, 2022.

\bibitem{Mao2018}
Y.~Mao, B.~Clerckx, and V.~O. Li, ``{Rate-Splitting Multiple Access for Downlink Communication Systems: {B}ridging, Generalizing, and Outperforming {SDMA and NOMA}},'' {\em Journal on Wireless Communications and Networking}, no.~133 (2018), 2018.

\bibitem{10741240}
J.~Xu and B.~Clerckx, ``{Max-Min Fairness and {PHY}-Layer Design of Uplink MIMO Rate-Splitting Multiple Access with Finite Blocklength},'' {\em IEEE Transactions on Communications}, pp.~1--1, 2024.

\bibitem{9831048}
Y.~Xu, Y.~Mao, O.~Dizdar, and B.~Clerckx, ``{Rate-Splitting Multiple Access With Finite Blocklength for Short-Packet and Low-Latency Downlink Communications},'' {\em IEEE Transactions on Vehicular Technology}, vol.~71, no.~11, pp.~12333--12337, 2022.

\bibitem{clerckx2023primer}
B.~Clerckx, Y.~Mao, E.~A. Jorswieck, J.~Yuan, D.~J. Love, E.~Erkip, and D.~Niyato, ``{A Primer on Rate-Splitting Multiple Access: Tutorial, Myths, and Frequently Asked Questions},'' {\em IEEE Journal on Selected Areas in Communications}, vol.~41, no.~5, pp.~1265--1308, 2023.

\bibitem{9970313}
J.~Xu, O.~Dizdar, and B.~Clerckx, ``{Rate-Splitting Multiple Access for Short-Packet Uplink Communications: A Finite Blocklength Analysis},'' {\em IEEE Communications Letters}, vol.~27, no.~2, pp.~517--521, 2023.

\bibitem{9991090}
Y.~Xu, Y.~Mao, O.~Dizdar, and B.~Clerckx, ``{Max-Min Fairness of Rate-Splitting Multiple Access With Finite Blocklength Communications},'' {\em IEEE Transactions on Vehicular Technology}, vol.~72, no.~5, pp.~6816--6821, 2023.

\bibitem{liu2023network}
Y.~Liu, B.~Clerckx, and P.~Popovski, ``{Network Slicing for eMBB, URLLC, and mMTC: An Uplink Rate-Splitting Multiple Access Approach},'' {\em IEEE Transactions on Wireless Communications}, 2023.

\bibitem{10243579}
M.~Diamanti, C.~Pelekis, E.~E. Tsiropoulou, and S.~Papavassiliou, ``{Delay Minimization for Rate-Splitting Multiple Access-Based Multi-Server MEC Offloading},'' {\em IEEE/ACM Transactions on Networking}, vol.~32, no.~2, pp.~1035--1047, 2024.

\bibitem{xiao2023delay}
F.~Xiao, P.~Chen, H.~Wu, Y.~Mao, and H.~Liu, ``{Delay Minimization Using Hybrid RSMA-TDMA for Mobile Edge Computing},'' {\em Electronics}, vol.~12, no.~11, p.~2550, 2023.

\bibitem{10032159}
P.~Chen, H.~Liu, Y.~Ye, L.~Yang, K.~J. Kim, and T.~A. Tsiftsis, ``{Rate-Splitting Multiple Access Aided Mobile Edge Computing With Randomly Deployed Users},'' {\em IEEE Journal on Selected Areas in Communications}, vol.~41, no.~5, pp.~1549--1565, 2023.

\bibitem{8259329}
Y.~Hu, M.~Serror, K.~Wehrle, and J.~Gross, ``{Finite Blocklength Performance of Cooperative Multi-Terminal Wireless Industrial Networks},'' {\em IEEE Transactions on Vehicular Technology}, vol.~67, no.~7, pp.~5778--5792, 2018.

\bibitem{5452208}
Y.~Polyanskiy, H.~V. Poor, and S.~Verdu, ``{Channel Coding Rate in the Finite Blocklength Regime},'' {\em IEEE Transactions on Information Theory}, vol.~56, no.~5, pp.~2307--2359, 2010.

\bibitem{7156144}
W.~Yang, G.~Caire, G.~Durisi, and Y.~Polyanskiy, ``{Optimum Power Control at Finite Blocklength},'' {\em IEEE Transactions on Information Theory}, vol.~61, pp.~4598--4615, Sep. 2015.

\bibitem{ozcan2013throughput}
G.~Ozcan and M.~C. Gursoy, ``{Throughput of Cognitive Radio Systems with Finite Blocklength Codes},'' {\em IEEE Journal on Selected Areas in Communications}, vol.~31, no.~11, pp.~2541--2554, 2013.

\bibitem{xu2016energy}
S.~Xu, T.-H. Chang, S.-C. Lin, C.~Shen, and G.~Zhu, ``{Energy-Efficient Packet Scheduling with Finite Blocklength Codes: Convexity Analysis and Efficient Algorithms},'' {\em IEEE Transactions on Wireless Communications}, vol.~15, no.~8, pp.~5527--5540, 2016.

\bibitem{hu2016blocklength}
Y.~Hu, A.~Schmeink, and J.~Gross, ``{Blocklength-Limited Performance of Relaying Under Quasi-Static Rayleigh Channels},'' {\em IEEE Transactions on Wireless Communications}, vol.~15, no.~7, pp.~4548--4558, 2016.

\bibitem{yang2014quasi}
W.~Yang, G.~Durisi, T.~Koch, and Y.~Polyanskiy, ``{Quasi-Static Multiple-Antenna Fading Channels at Finite Blocklength},'' {\em IEEE Transactions on Information Theory}, vol.~60, no.~7, pp.~4232--4265, 2014.

\bibitem{scarlett2016dispersion}
J.~Scarlett, V.~Y. Tan, and G.~Durisi, ``{The Dispersion of Nearest-Neighbor Decoding for Additive Non-Gaussian Channels},'' {\em IEEE Transactions on Information Theory}, vol.~63, no.~1, pp.~81--92, 2016.

\bibitem{schiessl2018delay}
S.~Schiessl, H.~Al-Zubaidy, M.~Skoglund, and J.~Gross, ``{Delay Performance of Wireless Communications with Imperfect CSI and Finite-Length Coding},'' {\em IEEE Transactions on Communications}, vol.~66, no.~12, pp.~6527--6541, 2018.

\bibitem{hu2018optimal}
Y.~Hu, M.~Ozmen, M.~C. Gursoy, and A.~Schmeink, ``{Optimal Power Allocation for QoS-Constrained Downlink Multi-User Networks in The Finite Blocklength Regime},'' {\em IEEE Transactions on Wireless Communications}, vol.~17, no.~9, pp.~5827--5840, 2018.

\bibitem{zhu2022energy}
Y.~Zhu, Y.~Hu, A.~Schmeink, and J.~Gross, ``{Energy Minimization of Mobile Edge Computing Networks with HARQ in The Finite Blocklength Regime},'' {\em IEEE Transactions on Wireless Communications}, vol.~21, no.~9, pp.~7105--7120, 2022.

\bibitem{lai2024short}
X.~Lai, T.~Wu, C.~Pan, L.~Mai, and A.~Nallanathan, ``{Short-Packet Edge Computing Networks With Execution Uncertainty},'' {\em IEEE Transactions on Green Communications and Networking}, 2024.

\bibitem{liu2018offloading}
J.~Liu and Q.~Zhang, ``{Offloading Schemes in Mobile Edge Computing for Ultra-Reliable Low Latency Communications},'' {\em Ieee Access}, vol.~6, pp.~12825--12837, 2018.

\bibitem{wang2017joint}
F.~Wang, J.~Xu, X.~Wang, and S.~Cui, ``{Joint Offloading and Computing Optimization in Wireless Powered Mobile-Edge Computing Systems},'' {\em IEEE transactions on wireless communications}, vol.~17, no.~3, pp.~1784--1797, 2017.

\bibitem{grant2009cvx}
M.~Grant, S.~Boyd, and Y.~Ye, ``{CVX: Matlab software for disciplined convex programming},'' 2009.

\end{thebibliography}

\end{document}